\documentclass[%
 aip,
 amsmath,amssymb,
 reprint,%
]{revtex4-1}

\usepackage{graphicx}
\usepackage{dcolumn}
\usepackage{bm}
\usepackage{amsmath}
\usepackage{hyperref}

\usepackage[utf8]{inputenc}
\usepackage[T1]{fontenc}
\usepackage{mathptmx}
\usepackage{etoolbox}
\usepackage{xcolor}

\makeatletter
\def\@email#1#2{%
 \endgroup
 \patchcmd{\titleblock@produce}
  {\frontmatter@RRAPformat}
  {\frontmatter@RRAPformat{\produce@RRAP{*#1\href{mailto:#2}{#2}}}\frontmatter@RRAPformat}
  {}{}
}%
\makeatother
\begin{document}

\preprint{AIP/123-QED}

\title{Enhancement of the Third Harmonic Generation Efficiency of ITO nanolayers coupled to Tamm Plasmon Polaritons}

\author{Tornike Shubitidze}
\affiliation{Department of Electrical \& Computer Engineering and Photonics Center, Boston University, 8 Saint Mary's Street, Boston, 02215, MA, USA}

\author{Smridhi Chawla}
\affiliation{Department of Physics, Boston University, 
590 Commonwealth Avenue, Boston,02215, MA, USA}

\author{Luca Dal Negro$^{*}$}\email{dalnegro@bu.edu}
\affiliation{Department of Electrical \& Computer Engineering and Photonics Center, Boston University, 8 Saint Mary's Street, Boston, 02215, MA, USA}
\affiliation{Department of Physics, Boston University, 
590 Commonwealth Avenue, Boston,02215, MA, USA}
\affiliation{Division of Materials Science \&  Engineering, Boston University, 15 St. Mary’s street, Brookline, 02446, MA, USA}

\date{\today}

\begin{abstract}
We study the enhancement of third harmonic generation in indium tin oxide nanolayers coupled to Tamm plasmon polaritons (TPPs). The TPPs are excited at the interface between a thin gold mirror and a silicon dioxide/silicon nitride (SiO$_2$/Si$_3$N$_4$) distributed Bragg reflector with a 30nm-thick indium tin oxide (ITO) nanolayer embedded inside the topmost dielectric layer under the metal mirror. This ITO nanolayer exhibits epsilon-near-zero (ENZ) behavior at near-infrared wavelengths. By tuning the angle of incidence and the TPP resonance conditions, we achieve sub-wavelength confinement of the electromagnetic field, resulting in a 8$\times$ enhancement of the nonlinear optical response of the structure compared to the isolated ITO nanolayer at its optimal ENZ condition. We further investigate the dependence of the THG signal on the incident angle and sample orientation, confirming that the enhancement is driven by the excitation of the TPP mode with a characteristic asymmetric  behavior. Numerical simulations of local field factors based on the transfer matrix method (TMM) fully support our findings. 
Our study demonstrates that the TPP-ENZ platform offers a versatile and highly efficient approach to enhancing nonlinear optical processes, with potential applications in frequency conversion, optical signal processing, and the development of more efficient nonlinear photonic devices. 
\end{abstract}

\maketitle

\section{Introduction}\label{intro}

The pursuit for highly nonlinear nanostructures marks a major frontier in photonics research, holding the potential for groundbreaking advancements in a wide range of applications such as telecommunications, signal processing, and quantum information technology\cite{Dutt2024}. In particular third-harmonic generation (THG) processes, wherein three photons are combined to generate a single photon at triple the fundamental frequency, are instrumental in extending the capabilities of integrated optical systems\cite{NonlinearOptics_Boydbook}. However, the efficiency of THG in traditional material platforms is inherently limited by the small value of the nonlinear susceptibility in conventional solid-state materials. As such, there is currently a compelling need to explore novel materials and nanoscale structures with enhanced efficiency and versatility. Recently, materials characterized by extremely low dielectric permittivity, dubbed epsilon-near-zero (ENZ) media, have emerged as promising candidates for achieving exceptionally efficient nonlinear optical effects on a solid-state platform at the nanoscale\cite{Reshef2019_ENZ_Review}. They provide remarkable enhancement of nonlinear optical processes such as harmonic generation, frequency mixing and conversion, electro-optical modulation, and extremely large light-induced refractive index variations in the nonperturbative regime \cite{Liberal2017, Reshef2019_ENZ_Review, Niu2018, Wu2021, Shubitidze2023, Khurgin2021,Tamashevich2023, Ciattoni2016, Guo2016, Bohn2021, Vezzoli2018, Capretti2015_THG, Capretti2015_SHG}. Specifically, a number of impressive ENZ-driven demonstrations of enhanced nonlinear phenomena and devices have been established based on indium tin oxide (ITO) and
indium doped cadmium oxide (CdO:In) nanolayers, including enhanced second- and third-harmonic generation (SHG/THG) \cite{Capretti2015_SHG,Capretti2015_THG,Yang2019,}, sub-picosecond all-optical modulation \cite{Guo2016}, high-efficiency optical time reversal \cite{Vezzoli2018}, as well as diffraction effects unconstrained by the pump bandwidth for time-varying and spatiotemporal photonic devices \cite{Tirole2022}. These phenomena, which are boosted by sub-wavelength field confinement in nanostructures across the ENZ spectral region, provide novel opportunities for classical and quantum information technologies, materials analysis, optical spectroscopy, and non-demolition photon detection schemes on the scalable silicon photonics platform. Moreover, ENZ structures based on ITO and CdO:In materials feature a wide tunability of the linear optical dispersion with a zero permittivity wavelength ENZ that extend from the near-infrared (NIR) to mid-infrared spectral range \cite{Wang2015,Runnerstrom2017,Britton2022,KellyK2019}, limited primarily by the optical losses at the ENZ wavelength. 

\begin{figure*}
\includegraphics[width=\textwidth]{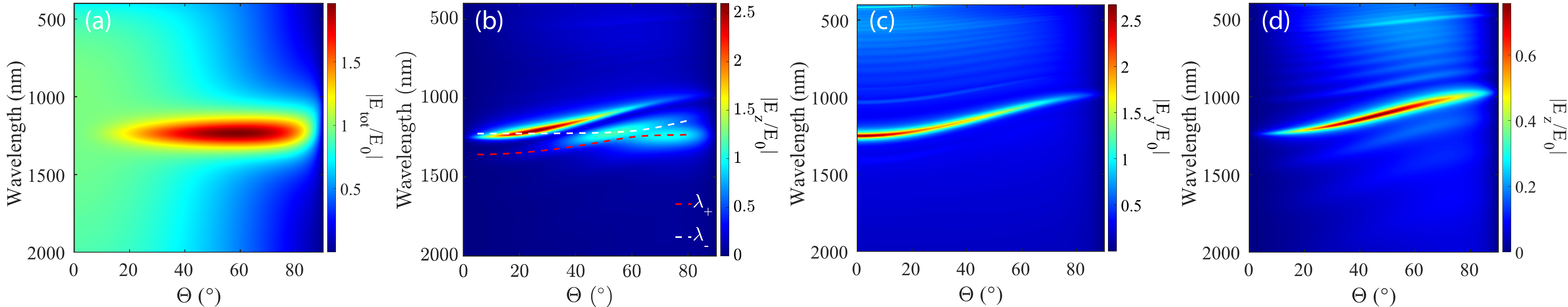}
\caption{\label{fig:ITO and TPP theory} Electric field enhancement in ITO nanolayers and nonlinear Tamm plasmon polariton structures as a function of wavelength and incident angle, $\Theta$ with TM polarization normalized by the incident field amplitude, $E_0$. (a) Total electric field enhancement of 30nm thick ITO film.(b-c) $E_z$ and $E_y$ field amplitudes within the embedded ITO nanolayer of the nonlinear TPP structure, respectively. The dashed red and white lines in panel (b) are the corresponding upper, $\lambda_+$, and lower branches, $\lambda_-$, of coupled mode wavelengths calculated from TCMT. (d)$E_z$ field amplitudes within a standard TPP structure with no embedded ENZ media.}
\end{figure*}

In parallel, Tamm plasmon polaritons (TPPs) have garnered attention for their ability to confine and enhance electromagnetic fields in surface waves at the nanoscale. Photonic TPP structures are composed of a distributed Bragg reflector (DBR) terminated by a thin metal film \cite{Kaliteevski2007}. Unlike traditional surface plasmon-polaritons (SPP), TPP surface states are excited within the optical gap of the DBR with strong field enhancement located at the interface between the metal and DBR. TPPs have a number of desirable optical properties that are well-suited for integrated nonlinear nano-optics. Specifically, their dispersion curve is entirely above the free photon light-line enabling broadband free space optical excitation in a compact structure. Moreover, TPP modes can be excited using either TE or TM polarized light at any incident angle. Finally, it has been shown that TPP modes exhibit an ultrafast response with measured mode lifetimes of the order of 20fs, significantly outpacing traditional SPPs at metal-dielectric interfaces\cite{Afinogenov2016}. Importantly, optimally designed dielectric photonic structures supporting optical Tamm plasmon polariton states give rise to enhanced electromagnetic field intensities at the nanoscale compared to conventional SPPs, significantly increasing the efficiency of nonlinear optical processes and even resulting in the strong coupling between cavity polaritons and Tamm surface plasmons \cite{Kaliteevski2009, Symonds2009,Grossmann2011}. 
Recently, nonlinear TPP structures composed of silicon dioxide/silicon nitride (SiO$_2$/SiN) with embedded ITO nanolayers fabricated by radio-frequency magnetron sputtering resulted in exceptional refractive index modulation $\Delta n \approx 2$ under weak pumping condition \cite{Shubitidze2023}.
Building upon these results, in our work, we investigate the enhancement of the THG efficiency in TPP structures with embedded ITO nanolayers. In particular, here we demonstrate a 8$\times$ enhancement in THG measured efficiency with respect to an isolated ITO nanolayer excited at the ENZ condition (i.e. minimal complex permittivity and optimal incidence angle) driven by the uniform enhancement of electric near-fields on the nanoscale. Our findings underscore the potential of ITO-integrated Tamm nanostructures for a wide range of photonic applications such as nonlinear spectroscopy, optical sensing, and quantum information technologies.

\section{Electric field enhancement in TPP-coupled ENZ nanolayers}\label{Results}
The investigated structures have been designed using the rigorous transfer matrix multilayer (TMM) approach to compute the electric field distributions inside the ITO nanolayers as a function of the angle of plane wave excitation under TE and TM polarization. The details of the full-vector internal field computation method, which follows references \cite{dal2003light,born2013principles}, are provided in our supplementary material.
Here we discuss the results for a 30nm thick ENZ ITO nanolayer on a silica substrate, as well as a multilayered structure supporting TPP states in the presence and in the absence of an embedded ITO nanolayer. All calculations have been performed with measured dielectric functions obtained from ref. \cite{Shubitidze2023}. In Figure \ref{fig:ITO and TPP theory} (a) we show the total electric field amplitude inside a 30nm ITO nanolayer on a fused silica substrate. The ITO nanolayer features an ENZ wavelength (i.e the wavelength at which the real part of permitivitty vanishes) at 1250nm with an imaginary permittivity  $\approx 0.4$. We found that the total electric field within the ITO nanolayer is maximized at a wavelength $\approx$1230nm for an incident angle $\approx60^{\circ}$. The wavelength of maximum electric field enhancement is therefore slightly blue shifted (by approximately 20nm) from the ENZ wavelength of the layer due to the intrinsic losses of the ITO material \cite{Capretti2015_SHG,Capretti2015_THG}. Moreover, the total electric field enhancement in the ITO nanolayer is dominated by the longitudinal $E_z$ field component, as previously reported \cite{Capretti2015_SHG,Capretti2015_THG}. In fact, we found that the transverse $E_y$ component, which is parallel to interface of the layer, only contributes at angles smaller than 40$^\circ$ with a reduced amplitude (see supplementary material). This sizeable enhancement of the longitudinal field component has been exploited to boost the efficiency of nonlinear processes in the proximity of the ENZ wavelength of ITO \cite{Capretti2015_THG,Reshef2019_ENZ_Review,Reshef2017}. In addition, ITO thin films benefit from material transparency in the UV-VIS wavelength range that enables the efficient extraction of the nonlinear harmonic generation when excited around their ENZ wavelength. However, in a single nanolayer of ENZ material the efficiency of nonlinear emission processes is still too small for any practical use in integrated optical devices \cite{Reshef2019_ENZ_Review}. 

In order to overcome these limitations, we propose to boost the field enhancement in ITO nanolayers using photonic structures with designed TPP states that are resonant around the ENZ response of the ITO nonlinear material. In particular, we consider structures consisting of 7-period distributed Bragg mirrors (DBR) composed of SiO$_2$/Si$_3$N$_4$ dielectric layers and terminated by a 24nm-thick gold mirror. Moreover, a 30nm-thick ITO nanolayer is embedded within the topmost Si$_3$N$_4$ layer directly underneath the gold mirror, at the location of the strongest electric field. We design the resonance frequency of the TPP structure in order to spectrally overlap the ENZ wavelength range of the ITO nanolayer and achieve maximum coupling and field enhancement in the spectral region with the strongest nonlinear response. By embedding the ITO film within the dielectric Si$_3$N$_4$ layer under the gold mirror, we ensure maximum field localization within the nonlinear ITO nanolayer \cite{Shubitidze2023}. In Figure \ref{fig:ITO and TPP theory}(b-c) we plot the $E_z$ and $E_y$ field amplitudes within the ITO nanolayer as a function of incident wavelength and angle for the coupled ITO-TPP structure. We notice that, in contrast to the case of the bare ITO film, the coupled structure shows a significant field enhancement for both components $E_{y,z}>2.5$ with a characteristic parabolic dispersion of surface TPP states \cite{Shubitidze2023}. In stark contrast, when no ITO nanolayer is present in the TPP structure, the corresponding $E_z$ field component is attenuated as shown in Figure \ref{fig:ITO and TPP theory}(d). Furthermore, in Figure \ref{fig:ITO and TPP theory}(b) at an angle of 60$^\circ$ we observe a characteristic hybridization behavior of the internal field in the spectral region corresponding to the largest intensity enhancement of the bare ITO nanolayer, previously shown in Figure \ref{fig:ITO and TPP theory}(a). This hybridization gives rise to a frequency splitting behavior that unveils the coupling between a photonic Fabry-Perot resonance of the TPP structure (at an oblique angle) and the ENZ mode of the ITO nanolayer. This coupling behavior can be modeled using Temporal Coupled Mode Theory (TCMT)\cite{Fan:03,haus1984waves,Suh2004}, where the energy positions and linewidths of the individual modes are retrieved from the optical absorption spectra of the TPP structure and the ITO nanolayer. A splitting energy of 0.011eV is obtained by fitting the level repulsion of the two coupled modes. The calculated resonant frequencies are represented as dashed lines in Figure 1(b), and their spectral positions agree closely with simulation results unveiling the nature of the coupling mechanism between ENZ materials and photonic modes. 
 This photonic coupling of the optical modes in the proposed nonlinear TPP structure is further investigated in our supplemental material where we address the impact of the optical losses in the ENZ media on the achievement of strong coupling regime using designed TPP structures.

\begin{figure}
\includegraphics[width=0.45\textwidth]{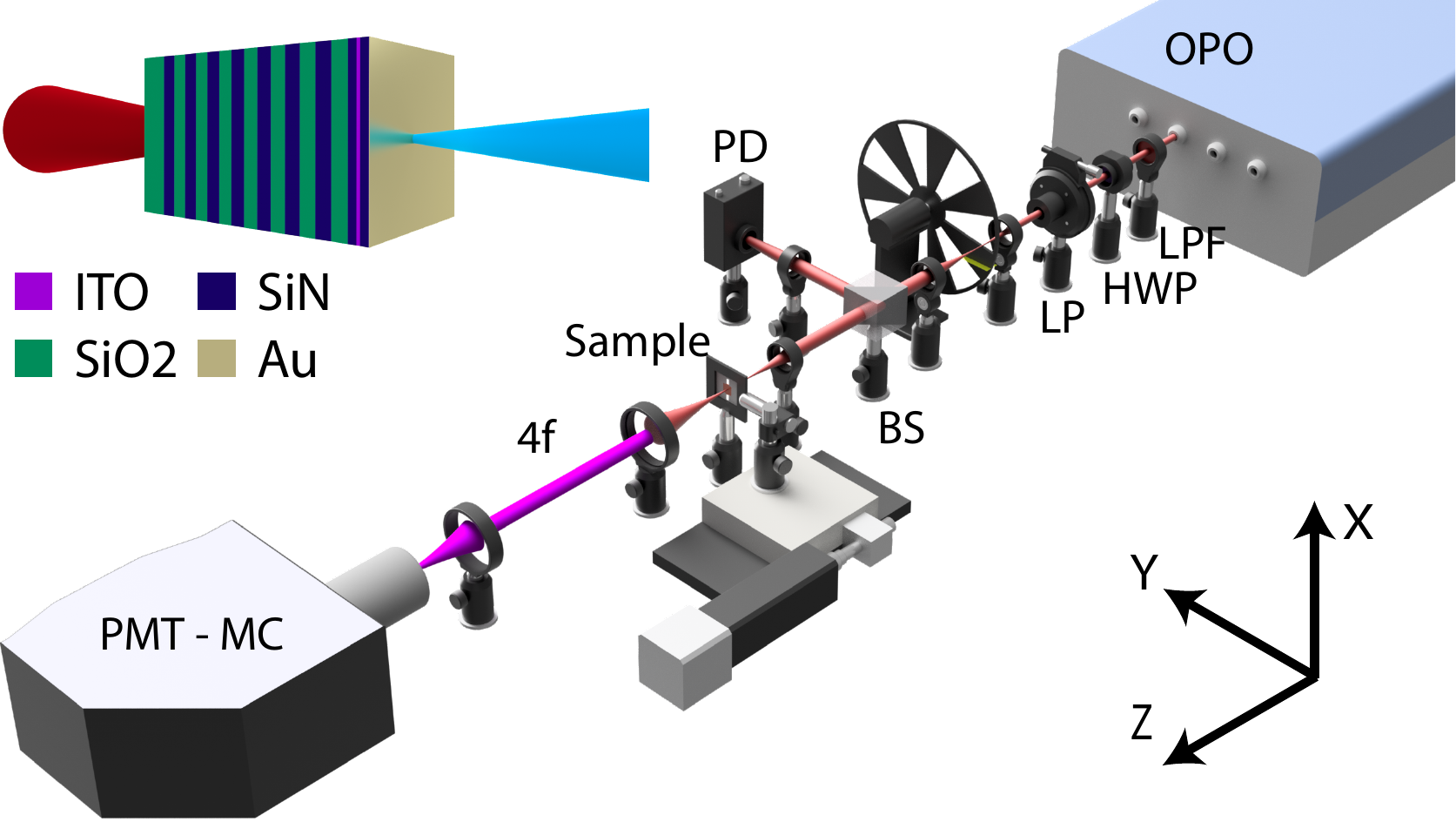}
\caption{Third-harmonic generation spectroscopy experimental setup. }\label{setup} 
\end{figure}

\begin{figure*}[t]
\includegraphics[width=\textwidth]{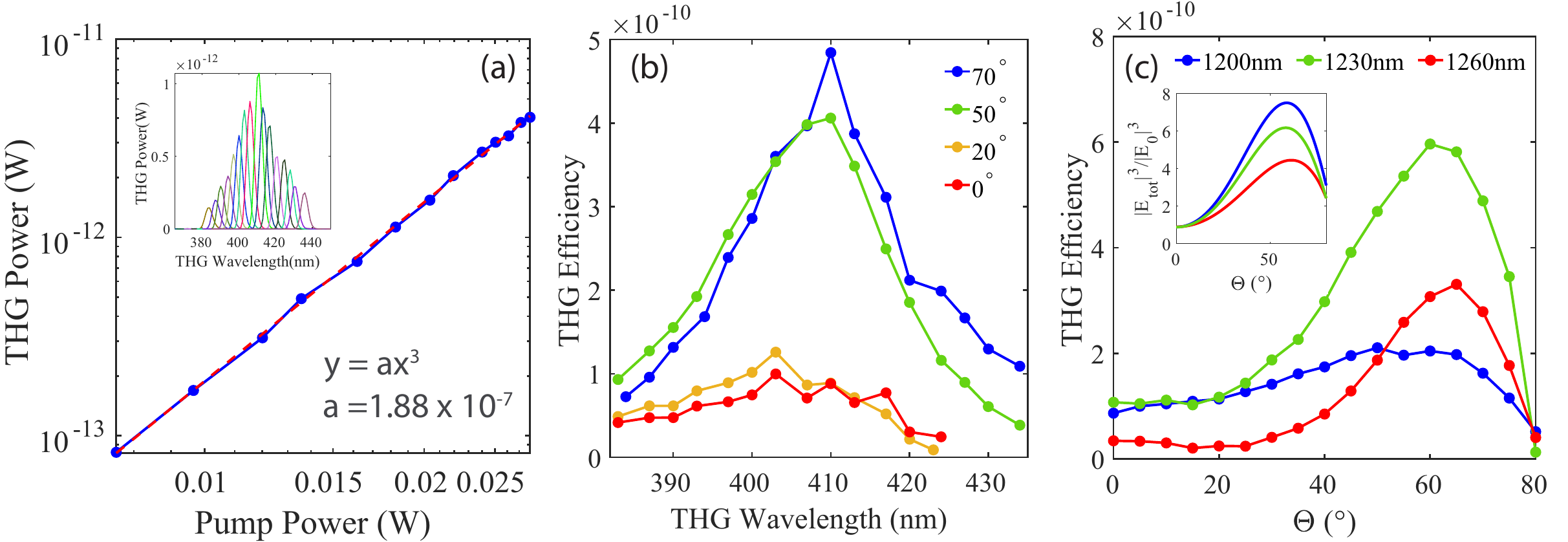}
\caption{\label{fig:30nmITO_THG} Third harmonic generation characterization of 30nm thick ITO films. (a) Third harmonic generation measurements (blue dots) as a function of average pump power at $\lambda_{pump} = 1230nm$ along with a cubic fit (dashed red). The inset shows representative THG spectra at different pump wavelengths from 1150-1300nm. (b) THG efficiency measurements at various angles of incidence as a function of pump wavelength. (c) THG efficiency measurements as a function of incident angle at various pump wavelengths. The inset shows the theoretical pump enhancement cubed of 30nm ITO films at each respective wavelength according to the TMM caculations in Figure \ref{fig:ITO and TPP theory}(a).}
\end{figure*}

\section{Experimental Setup for nonlinear characterization}
The ITO nanolayers and nonlinear TPP structures were excited using the setup sketched in Figure \ref{setup}. Specifically, we use a Spectra-Physics Inspire optical parametric oscillator (OPO) pumped by an 80 MHz repetition rate Mai Tai HP Ti:Sapphire tunable laser with 150 fs pulses with TEM00 idler output at near-infrared wavelengths in transmission mode. The laser power was attenuated by rotating a half-wave plate (HWP) positioned before a linear polarizer (LP) oriented for TM polarization. A long pass filter (LPF) was used directly after the OPO in order to extinguish the remaining second harmonic beam within the OPO cavity. Two converging lenses act as a telescope that both magnified the idler output and focused the beam to a small spot size at the plane of a mechanical chopper. An 80:20 ultrafast beam splitter (BS) was used to monitor the average power of the NIR beam and the transmitted idler beam was focused down onto the sample by a 50mm focal length plano-convex lens. The THG signals are collected and focused down onto a Cornerstone 260 monochromator (MC) with a 4f system and detected by a lock-in amplifier coupled to a low-light photomultiplier tube (PMT, Newport Oriel 77348). All measurements were taken in a completely darkened environment. A calibration procedure was performed in order to accurately estimate the efficiencies of the THG process as a ratio between the generated THG intensity and the initial pump power intensity. The calibration procedure is as follows: a fiber coupled 405nm laser (Thorlabs) was greatly attenuated and its intensity was measured directly before the monochromator entrance with a calibrated photodiode and later with the MC-PMT setup. This procedure relates the signal measured by the PMT (in units of Volts) to the average optical power generated at the sample plane (in units of Watts). This experimental setup is capable of measuring harmonic generation efficiencies as low as $10^{-13}$
\section{THG from bare ITO nanolayers}
The 30nm-thick ITO nanolayers were fabricated by RF magnetron sputtered from an ITO (In$_2$O$_3$/SnO$_2$ 90/10 wt.\%) target on transparent fused silica glass. The samples were fabricated at room temperature with an RF power of 150W in a pure Ar environment. The as-deposited ITO films were then vacuum-annealed at 400$^\circ$C for 1 hour in order to introduce oxygen vacancies, increasing the electron density and mobility leading to a blue-shift of the ENZ crossing point to $\approx$ 1250nm. The samples were analyzed using broad band variable angle spectroscopic ellipsometry (VASE) and normal incidence transmission measurements. Further details of the fabrication and linear optical property characterization of these samples can be found in reference \cite{Shubitidze2023}. In Figure \ref{fig:30nmITO_THG} we summarize the results on the THG optical response of the fabricated 30nm thick ITO nanolayers. In particular, Figure \ref{fig:30nmITO_THG}(a) measurements performed at a pump wavelength of 1230nm as a function of incident average pump power (blue dots) demonstrate the cubic scaling of the THG intensity. The dashed red line is a cubic fit showing excellent agreement with measurement and confirming the signal is nonlinearly generated third-harmonic. Subsequently, we recorded the TH efficiencies as a function pump wavelength around the ENZ condition, from 1150nm-1300nm with 10nm spacing, for various fixed angles of incidence as shown in Figure \ref{fig:30nmITO_THG}(b). We kept the incident pump power constant at an average of 34mW corresponding to a peak intensity of 0.45 GW/cm$^2$.  Our measurements show a clear spectral dependence of the THG conversion efficiency at oblique incident angles greater than 20$^\circ$ near the ENZ condition with a maximum efficiency of $6\times10^{-10}$ measured at a pump wavelength of 1230nm. Given the deep sub-wavelength thickness of this film and the exceptionally low peak pump intensities used in this work, the measured THG efficiencies are over 50$\times$ stronger than those previously reported in the literature, when considering comparable peak pump intensities\cite{Luk2015}. This underscores the critical importance of engineering high-quality ENZ materials for enhancing third harmonic generation and advancing nonlinear optical applications. At normal and 20$^\circ$ incident angle we find insignificant spectral dependence of the THG efficiency consistent with our theoretical findings based on field enhancement. Moreover, we report THG measurements as a function of incident angle at pump wavelengths at and around the ENZ condition using pump wavelengths of 1200nm, 1230nm and 1260nm in Figure \ref{fig:30nmITO_THG}(c). From these data, we find a clear angular dependence of the THG signal with an enhancement of 7$\times$ occurring at 65$^\circ$ incident angle and 1230nm pump wavelength with respect to normal incidence THG measurements.
In order to elucidate the mechanism behind the angular and spectral dependence of the THG intensity behavior of ENZ nanolayers we investigate the local field factors (LFF) within the medium. The LFF quantifies the magnitude ratio between the free-space field and the local field within the medium\cite{novotny_hecht_2012}. In the context of optical-harmonic generation in the dipole approximation, the far-field of the emitted $n$th harmonic radiation $E(\omega n)$ depends on the free-space pump field $E(\omega)$ as $E(\omega n) = L^n(\omega)\dot\chi^{(n)}E^n(\omega)$ where $\chi^{(n)}$ is the nth order nonlinear susceptibility, L($\omega$) is the LFF for the fundamental pump wavelength\cite{novotny_hecht_2012}. Intensities of second- and third-harmonic signals have shown to be enhanced as the fundamental LFF is increased in the presence of localized electric field modes such as surface and localized plasmons \cite{Wang2013_Fano}, or microcavity modes in photonic crystals\cite{Martemyanov2004}. Using the LFFs at the fundamental pump wavelengths investigated in this work, we calculate the THG enhancement  for our ITO nanolayers as a function of incident angle. The results are plotted in the inset of Figure \ref{fig:30nmITO_THG}(c) and are in good agreement with our measurements yielding a THG efficiency enhancement of $\approx$7$\times$ at an incident angle of 60$^\circ$ with respect to normal incidence.

\section{THG from Nonlinear TPP-ITO coupled structures}

\begin{figure*}
\includegraphics[width=\textwidth]{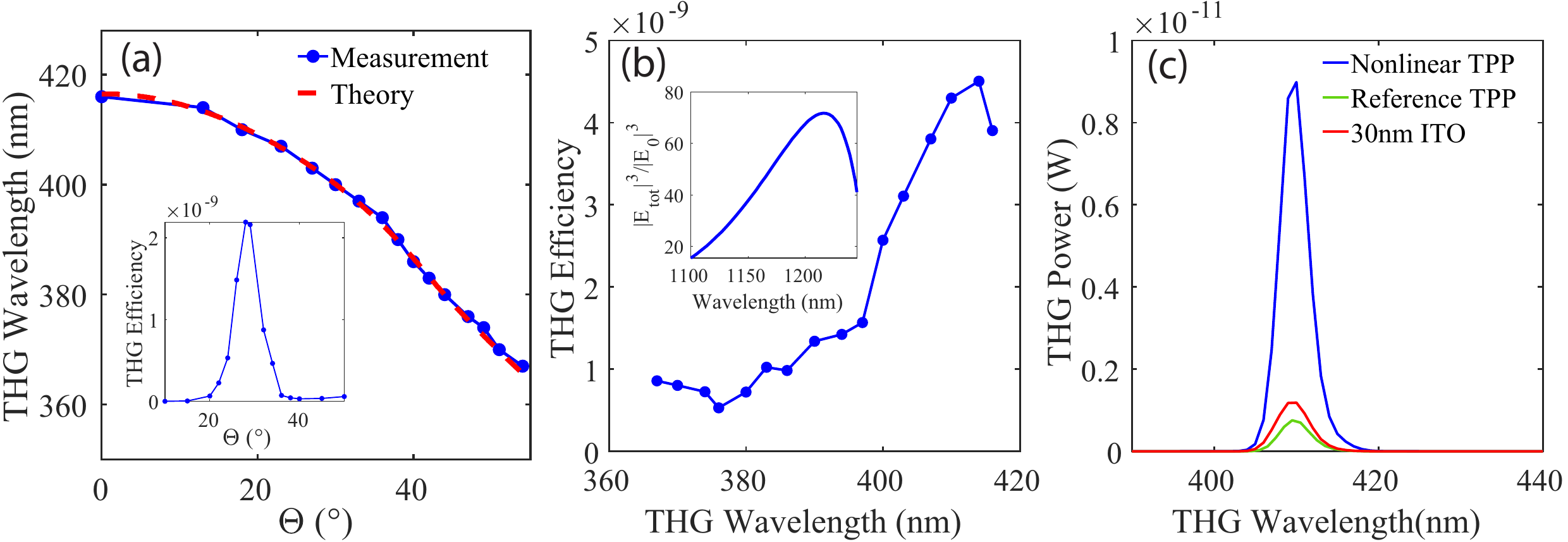}
\caption{\label{fig:1DTamm_THG} (a) Measured (blue dots) parabolic dispersion curve of the THG signal from nonlinear TPP structures along with the theoretical curve following the fundamental pump (dashed red). The inset shows the angular dependence of the THG efficiency at a pump wavelength of 1200nm.(b) Efficiency measurements of the THG intensity from nonlinear TPP structures as a function of pump wavelength. The inset shows the total field enhancement cubed within the ITO nanolayer as a function of wavelength calculated from TMM theory. (c) Direct comparison of the THG intensity from our nonlinear TPP sample, reference TPP sample and 30nm thick ITO nanolayer. The measurements were performed at a pump wavelength of 1230nm and the bare ITO film was oriented at an incidence angle of 65$^\circ$.}
\end{figure*}

In this section we demonstrate THG enhancement from the nonlinear TPP structures. In Figure \ref{fig:1DTamm_THG} we summarize our results. In particular, Figure \ref{fig:1DTamm_THG} (a) plots the measured (blue dots) THG angular dispersion curve as a function of THG emission wavelength, from 366nm to 416nm, and incident angle, from 0$^\circ$ to 60$^\circ$. These data perfectly follow the shift of the fundamental Tamm state from 1100nm to 1250nm (dashed red), proving the nonlinear coupling of the ITO nanolayer with the TPP structure. In fact, since the TPP resonance blue shifts as the angle of incidence increases, the THG resonance blue shifts accordingly. Furthermore, by fixing the pump wavelength at 1200nm and measuring the THG intensity as a function of incident angle, as shown in the inset of Figure \ref{fig:1DTamm_THG}(a), we demonstrate that no significant THG signals are present besides the fundamental TPP resonance. Moreover, in Figure \ref{fig:1DTamm_THG}(a) we measure the THG efficiency of the nonlinear TPP structure as function of pump wavelength using the same pump intensity. We find a clear spectral dependence of the THG intensity with a peak efficiency of $4.5 \times 10^{-9}$ at a pump wavelength of 1230nm, average incident power of 34 mW, and incident angle of 18$^{\circ}$. The measured efficiency of these nonlinear TPP structures, at the low peak pump intensities used in this work, corresponds to over two-orders of magnitude stronger THG emission than typical optical Tamm state structures\cite{Afinogenov2018}. Moreover, we find a 3$\times$ increase in THG efficiency from ITO nanolayers compared to the traditional Kretschmann geometries that are used to excite ENZ modes within ITO thin films for enhanced THG efficiency\cite{Luk2015}. A gradual decrease in the TH intensity as the angle of incidence is increased beyond 18$^\circ$ is observed. This behavior is explained by the gradual decrease of the total field enhancement within the nonlinear TPP structure as the angle of incidence is increased beyond $\approx 20^\circ$, thus blue shifting the resonant wavelength. The inset of Figure \ref{fig:1DTamm_THG}(b) shows the calculated maximum THG enhancement of our nonlinear TPP structure based on the fundamental LFF as a function of wavelength, and indicates a good agreement with the measurement. With respect to the maximum THG efficiency of the bare ITO nanolayer at a pump wavelength of 1230nm, we find a 8$\times$ increase in efficiency in the designed nonlinear TPP structure. This is consistent with the LFF calculations at resonance within the nonlinear TPP structure which is 2.2$\times$ greater than the LFF at the ENZ wavelength of the ITO nanolayer. Furthermore we measure the THG emission from a standard reference TPP structure without the embedded ITO nanolayer and found that it is $\approx$2$\times$ smaller compared to the one of a bare 30nm-thick ITO nanolayer, consistent with our previous measurements in reference \cite{Shubitidze2023}. The experimental results confirm that TPP structures with embedded ITO nanolayers exhibit a 8$\times$ enhancement in THG efficiency compared to bare ITO films at the fundamental ENZ wavelength of 1230nm. This enhancement is directly manifested due to the strong nanoscale localization of the electric field within the nonlinear TPP structures, driven by the strong coupling regime of the ENZ and photonic modes. 

\begin{figure}
\includegraphics[width=0.45\textwidth]{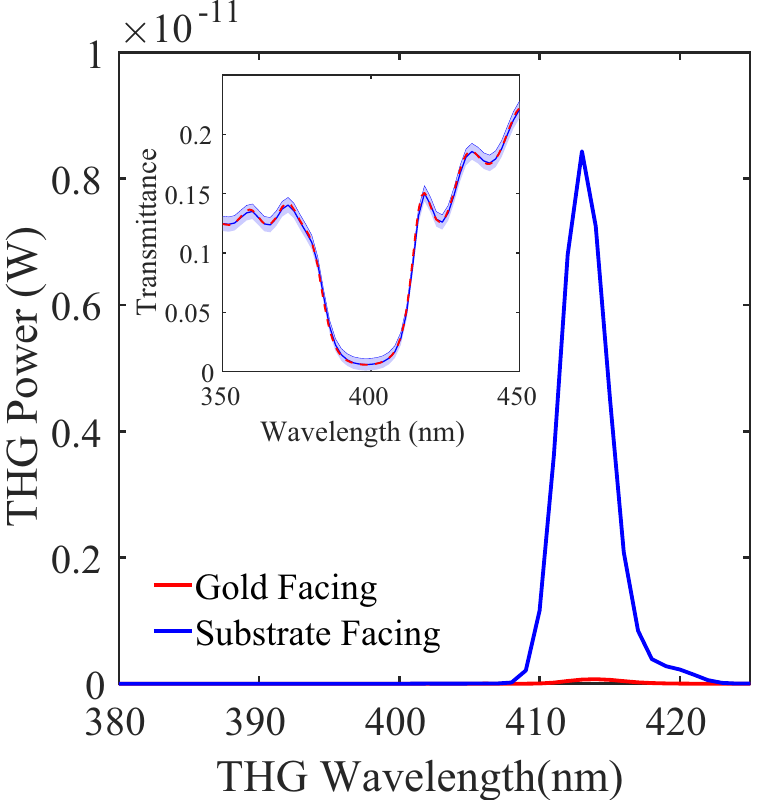}
\caption{\label{fig:1D Tamm Reciprocity} Asymmetric transmission behavior of the THG signal from nonlinear TPP structures. Measured THG power when the structure is pumped substrate (blue) or gold layer first (red). The inset shows the linear transmission measurement of the nonlinear TPP structure at normal incidence from 350nm-450nm in both excitation orientations. The blue shading corresponds to the error bars of the measurement.}
\end{figure}

Finally, we report on the another remarkable feature of nonlinear TPP structures, which give rise to a characteristic asymmetric behavior when exciting the sample from different directions  \cite{Afinogenov2018}. In Figure \ref{fig:1D Tamm Reciprocity} we demonstrate the asymmetric response of our TPP nonlinear sample by showing the emitted THG signal when the sample is oriented with either the gold layer or the substrate facing the incident beam. From Figure \ref{fig:1D Tamm Reciprocity} we find a 100X increase in THG signal in the substrate-first orientation, with negligible THG observed in the gold-facing configuration. To understand this behaviour, we measure the linear transmittance of the TPP sample, in both orientations, at the THG wavelength and plot the results in the inset of Figure \ref{fig:1D Tamm Reciprocity}. We found the presence of a secondary optical bandgap and a transmittance of $\approx0.5\%$ within the bandgap with no TPP states, thus preventing the escape of the generated THG signal through the DBR stack. In order for TPP states to exist in this wavelength range, the following condition must be met, $R_mR_{DBR} = 1$, where $R_m$ is the amplitude reflection coefficient for the wave incident from the metal layer and $R_{DBR}$ is the amplitude reflection coefficient of the DBR. We found that with a thickness of 24nm, around $\lambda = 400nm$ the reflection coefficient of the gold mirror does not meet this condition. Therefore, no transmission TPP state can exist in the secondary optical bandgap of our nonlinear TPP structures, rendering its nonlinear emission strongly asymmetric. 

\section{Conclusions}\label{conclusion}

In this work, by leveraging the unique properties of ITO ENZ nanolayers and TPP structures, we achieved a 8$\times$ increase in THG efficiency compared to a bare ITO film at the optimal ENZ pumping conditions. This enhancement results primarily from the uniform and nanoscale localization of the electric field in the TPP structure, manifesting the coupling regime between ENZ and photonic modes. Our experimental data are supported by full-vector electromagnetic field calculations of LFFs that show significant electric field enhancement in the ITO-coupled TPP structure for both $E_y$ and $E_z$ components. 
Furthermore, our findings demonstrate that the THG efficiency of the designed ITO-coupled TPP structure is 8$\times$ larger than in the ENZ enhanced efficiency of the bare ITO films with a peak efficiency of $4.5 \times 10^{-9}$ at a pump wavelength of 1230nm and average incident power of only 34 mW. Finally, we report on the remarkable asymmetric behavior of the investigated TPP structure with an observed a 100$\times$ increase in THG signal intensity when it was oriented with the substrate facing the incident beam. We believe that our findings underscore the potential of ITO-TPP strongly coupled systems, for a wide range of photonic applications such as optical sensing and spectroscopy, signal processing, and novel quantum detection modalities. 

\section{Supplemental}

The supporting simulation, and additional analysis are provided in the supplementary material.

\section{Acknowledgements} 
This research was sponsored by the U.S. Army Research Office and accomplished under award number W911NF2210110.

\bibliography{references_THG_Tamm}

\end{document}